\documentstyle[buckow]{article}

\def\l{\lambda}
\def\z{\zeta}
\def\s{\sigma}
\def\ve{\varepsilon}

\def\beq{\begin{equation}}
\def\eeq{\end{equation}}

\begin {document}
\def\titleline{
Quantum Geometry of the Universal Hypermultiplet
}
\def\authors{
Sergei V. Ketov~\footnote{On leave from the University of Kaiserslautern, 
Germany}}
\def\addresses{
Caltech-USC Center for Theoretical Physics\\  
Department of Physics and Astronomy \\
University of Southern California \\ 
Los Angeles, CA 90089-2535, USA \\
{\tt e-mail ketov@citusc.usc.edu} \\
}
\def\abstracttext{
The universal hypermultiplet moduli space metric in the type-IIA superstring theory compactified on 
a Calabi-Yau threefold is related to integrable systems. The instanton corrections in four dimensions 
arise due to multiple wrapping of BPS membranes and fivebranes around certain (supersymmetric) cycles of 
Calabi-Yau. The exact (non-perturbative) metrics can be calculated in the special cases of (i) the 
D-instantons (or the wrapped D2-branes) in the absence of fivebranes, and (ii) the fivebrane instantons 
with vanishing charges, in the absence of D-instantons. The solutions of the first type are governed
by the three-dimensional Toda equation, whereas the solutions of the second type are governed by 
 the particular Painlev\'e VI equation.
}
\large
\makefront


\let\du=\du                     


\def\a{\alpha}

\def\c{\chi}
\def\d{\delta}

\def\f{\phi}

\def\k{\kappa}
\def\l{\lambda}
\def\m{\mu}
\def\n{\nu}
\def\o{\omega}
\def\p{\pi}

\def\s{\sigma}
\def\t{\tau}

\def\z{\zeta}

\def\L{\Lambda}
\def\O{\Omega}


\def\ve{\varepsilon}

\def\vq{\vartheta}


\def\cc{{\cal C}}

\def\cm{{\cal M}}

\def\cy{{\cal Y}}





\def\bo{{\raise-.5ex\hbox{\large$\Box$}}}               
\def\pa{\partial}                                       
\def\TH{{\raise.2ex\hbox{$\displaystyle \bigodot$}\mskip-4.7mu \llap H \;}}
\def\face{{\raise.2ex\hbox{$\displaystyle \bigodot$}\mskip-2.2mu \llap {$\ddot
        \smile$}}}                                      


\def\sp#1{{}^{#1}}                              
   %
       %
\def\VEV#1{\left\langle #1\right\rangle}        
\def\abs#1{\left| #1\right|}                    
\def\leftrightarrowfill{$\mathsurround=0pt \mathord\leftarrow \mkern-6mu
        \cleaders\hbox{$\mkern-2mu \mathord- \mkern-2mu$}\hfill
        \mkern-6mu \mathord\rightarrow$}
\def\dvec#1{\vbox{\ialign{##\crcr
        \leftrightarrowfill\crcr\noalign{\kern-1pt\nointerlineskip}
        $\hfil\displaystyle{#1}\hfil$\crcr}}}           
\def\dt#1{{\buildrel {\hbox{\LARGE .}} \over {#1}}}     


\def\frac#1#2{{\textstyle{#1\over\vphantom2\smash{\raise.20ex
        \hbox{$\scriptstyle{#2}$}}}}}                   
\def\sfrac#1#2{{\vphantom1\smash{\lower.5ex\hbox{\small$#1$}}\over
        \vphantom1\smash{\raise.4ex\hbox{\small$#2$}}}} 
\def\bfrac#1#2{{\vphantom1\smash{\lower.5ex\hbox{$#1$}}\over
        \vphantom1\smash{\raise.3ex\hbox{$#2$}}}}       
\def\afrac#1#2{{\vphantom1\smash{\lower.5ex\hbox{$#1$}}\over#2}}    

\def\[{\lfloor{\hskip 0.35pt}\!\!\!\lceil}
\def\]{\rfloor{\hskip 0.35pt}\!\!\!\rceil}

\def\du#1#2{_{#1}{}^{#2}}

\def\fracm#1#2{\hbox{\large{${\frac{{#1}}{{#2}}}$}}}

\def\fracmm#1#2{{{#1}\over{#2}}}

\def\low#1{{\raise -3pt\hbox{${\hskip 0.75pt}\!_{#1}$}}}

\def\Dot#1{\buildrel{_{_{\hskip 0.01in}\bullet}}\over{#1}}
\def\dt#1{\Dot{#1}}


\newskip\humongous \humongous=0pt plus 1000pt minus 1000pt
\def\caja{\mathsurround=0pt}
\def\eqalign#1{\,\vcenter{\openup2\jot \caja
        \ialign{\strut \hfil$\displaystyle{##}$&$
        \displaystyle{{}##}$\hfil\crcr#1\crcr}}\,}
\newif\ifdtup


\def\ref#1{$\sp{#1)}$}

\def\pl#1#2#3{Phys.~Lett.~{\bf {#1}B} (19{#2}) #3}
\def\np#1#2#3{Nucl.~Phys.~{\bf B{#1}} (19{#2}) #3}

\def\cqg#1#2#3{Class.~and Quantum Grav.~{\bf {#1}} (19{#2}) #3}

\section{Introduction}

Non-perturbative contributions to the effective supergravity theory, originating from the type 
IIA string compactification  on a {\it Calabi-Yau} (CY) threefold $\cy$, are known to be due to the 
solitonic five-branes wrapped about the entire CY space and the supermembranes (D2-branes) wrapped 
about special Lagrangian (supersymmetric) three-cycles $\cc_3$ of $\cy$ \cite{bbs}. The supersymmetric 
cycles minimize volume in their homology class, while the corresponding wrapped brane configurations 
lead to the BPS states. Being solitonic (BPS) classical solutions to the higher dimensional 
(Euclidean) equations of motion, these wrapped branes are localized in the uncompactified (four) dimensions 
and thus can be identified with 4d instantons. The instanton actions are essentially given by 
the volumes of the cycles on which the branes are wrapped.
 
The compactification of the type-IIA superstring theory on  $\cy$ gives rise to the four-dimensional 
(4d) N=2 superstrings whose {\it Low-Energy Effective Action} (LEEA) is given by the 4d, N=2
supergravity coupled to N=2 vector supermultiplets and hypermultiplets. The hypermultiplet 
LEEA is most naturally described by the {\it Non-Linear Sigma-Model} (NLSM), 
whose scalar fields parametrize the quaternionic target space $\cm_H$ 
\cite{bw}. The instanton corrections to the LEEA due to the wrapped 
fivebranes and membranes can be easily identified and distinguished from each 
other in the semi-classical limit, since the fivebrane instanton corrections 
are organized by powers of $e^{-1/g^2_{\rm string}}$, whereas the membrane 
instanton corrections are given by powers of $e^{-1/g_{\rm string}}$, where 
$g_{\rm string}$ is the type-IIA superstring coupling constant \cite{w1}.
The vacuum expectation value of the four-dimensional dilaton field $\VEV{\f}$ in
 the compactified type-IIA superstring is simply related to the CY volume 
$V_{\rm CY}$ in M-theory, $V_{\rm CY}=e^{-2\VEV{\f}}$, so that the type-IIA 
superstring loop expansion amounts to the derivative expansion of the M-theory
action \cite{one}. Any CY compactification has the co-called {\it Universal Hypermultiplet} 
(UH) containing a dilaton, an axion, a complex RR-type pseudo-scalar and a Dirac 
dilatino. The target space of the universal hypermultiplet NLSM has to be an
 Einstein space with the {\it (Anti)Self-Dual} (ASD) Weyl tensor \cite{bw}. We  restrict ourselves 
to a calculation of the instanton corrections to the universal hypermultiplet NLSM metric by analyzing generic 
quaternionic deformations of the classical UH metric. We use the simple fact that the (anti)self-dual Weyl 
tensor already implies the integrable system of partial differential equations on the components of the 
UH moduli space metric. Additional simplifications arise due to the Einstein condition and the physically 
motivated isometries. The exact UH metric is supposed to be regular and complete ({\it cf.} Seiberg-Witten
theory --- see, e.g., ref.~\cite{swrev} for a review). 

\section{UH metric in string perturbation theory}

The LEEA of (tree) type-IIA superstrings in ten dimensions is given by the IIA supergravity.  
The universal (UH) sector of the 10d type-IIA supergravity compactified down to four dimensions is 
obtained by using the following {\it Ansatz} for the 10d metric: 
$$ ds^2_{10}=g_{mn}dx^mdx^n
=e^{-\f/2}ds^2_{\rm CY}+e^{3\f/2}g_{\m\n}dx^{\m}dx^{\n}~,\eqno(1)$$
while keeping only $SU(3)$ singlets in the internal CY indices and ignoring all CY complex moduli. 
In eq.~(1) $\f(x)$ stands for the 4d dilaton, $g_{\m\n}(x)$ is the spacetime
metric in four uncompactified dimensions, $\m,\n=0,1,2,3$, and 
$ds^2_{\rm CY}$ is the (K\"ahler and Ricci-flat) metric of the internal CY 
threefold $\cy$ in complex coordinates, 
$$ds^2_{\rm CY}=g_{i\bar{j}}(y,\bar{y})dy^id\bar{y}^{\bar{j}}~,\eqno(2)$$
where $i,j=1,2,3$. By definition, the CY threefold $\cy$ possesses the $(1,1)$ 
K\"ahler form $J$ and the holomorphic $(3,0)$ form $\O$. The universal 
hypermultipet (UH) unites the dilaton $\f$, the axion $D$ coming from 
dualizing the three-form field strength $H_3=dB_2$ of the NS-NS two-form $B_2$ 
in 4d, and the complex scalar $C$ representing the RR three-form $A_3$ with 
$A_{ijk}(x,y) =\sqrt{2}C(x)\O_{ijk}(y)$. When using a flat (or rigid) CY  with 
$$  g_{i\bar{j}}=\d_{i\bar{j}}\quad {\rm and}\quad \O_{ijk}=\ve_{ijk}~,
\eqno(3)$$
this yields the (Ferrara-Sabharwal) NLSM action in 4d \cite{fsh},
$$
S_{4}=-\fracmm{1}{\k^2_{4}}\int d^{4}x\sqrt{-g}\left[ -\fracm{1}{2}{\cal R}
 + (\pa_m\f)^2 +e^{2\f}\abs{\pa_{\m}C}^2 +e^{4\f}\left(\pa_{\m}D+
\fracm{i}{2}\bar{C}\dvec{\pa_{\m}}C\right)^2\right]~.\eqno(4)$$
where $H_3$ has been traded for the pseudoscalar $D$ via the Legendre transform.

The perturbative (one-loop) string corrections to the UH metric originate from 
the $(Riemann)^4$ terms in M-theory compactified on a  CY three-fold $\cy$ \cite{one}. These quantum 
corrections are known to be proportional to the CY Euler number $\c=2\left(h_{1,1}-h_{1,2}\right)$. In fact, 
the corrected metric is related to the classical UH metric by a local field redefinition
\cite{one}, so that the local UH geometry is unchanged in superstring perturbation theory.

\section{D-instantons and UH metric}

The classical (Ferrara-Sabharwal) NLSM metric describes the symmetric quaternionic space 
$SU(2,1)/SU(2)\times U(1)$. In particular, the $U(1)$ subgroup of the $SU(2)$ symmetry 
is given by the duality rotations $U_C(1)$ of the complex R-R pseudo-scalar $C$,
$$ C\to e^{ i\a} C~. \eqno(5)$$
These duality rotations are believed to be  exact in quantum theory \cite{bbs}, as we assume too.

As regards generic four-dimensional quaternionic manifolds (relevant for UH), they all
 have {\it Einstein-Weyl\/} geometry of {\it negative\/} scalar curvature \cite{bw}, 
$$ W^-_{abcd}=0~,\qquad R_{ab}=\fracm{\L}{2}g_{ab}~,\eqno(6)$$  
where  $W_{abcd}$ is the Weyl tensor and  $R_{ab}$ is the Ricci tensor for 
the metric $g_{ab}$. When using the {\it Ansatz} \cite{tod} 
$$ ds^2_{\rm Q}= \fracmm{P}{\o^2}\left[ e^u (dx^2+dy^2)+d\o^2\right]
+\fracmm{1}{P\o^2}(dt +\Theta_1)^2 \eqno(7)$$
for a generic quaternionic metric with an abelian isometry, 
it is straightforward to prove that the restrictions (6) on the metric (7)
 {\it precisely\/} amount to the 3d  Toda equation   
$$ u_{xx} + u_{yy}+(e^u)_{\o\o}=0~.\eqno(8)$$
The second potential $P$ of eq.~(7) is then given by \cite{tod} 
$$ P= \fracmm{1}{2\L}\left(\o u_{\o}-2\right),\eqno(9)$$
whereas the remaining one-form $\Theta_1$ obeys the linear equation \cite{tod}
$$ d\Theta_1= -P_x\,dy\wedge d\o-P_y\,d\o\wedge dx-e^u(P_{\o}+\fracm{2}{\o}P
+\fracm{2\L}{\o}P^2)dx\wedge dy~.\eqno(10)$$ 
In terms of the complex coordinate $\z=x+iy$, the 3d Toda equation (8) takes 
the form
$$ 4u_{\z\bar{\z}}+(e^u)_{\o\o}=0~.\eqno(11)$$

{\it Separable} solutions to the 3d Toda equation, having the form
$$ u (\z,\bar{\z},\o) = F(\z,\bar{\z}) + G(\o)~,\eqno(12)$$
are easily found to be \cite{din}
$$ e^u= \fracmm{4c^2(\o^2+2\o b\cos\a +b^2)}{(1+c^2\abs{\z}{}^2)^2}~~~,
\eqno(13)$$
where $(\a,b,c)$ are all constants. Eq.~(13) automatically possesses the rigid $U_C(1)$ 
symmetry with respect to the duality rotations $\z\to e^{i\a}\z$ of the complex RR-field 
$\z$.

The classical approximation corresponds to the conformal limit $\o\to\infty$ and 
$\abs{\z}\to\infty$, while keeping the ratio $\abs{\z}^2/\o$ finite. Then one easily finds 
that $P\to -\L^{-1}=const.>0$, whereas the metric based on eq.~(13) takes the form
$$ ds^2=\fracmm{1}{\l^2}\left(\abs{dC}^2+d\l^2\right)
+\fracmm{1}{\l^4}(dD+\Theta)^2~,\eqno(14)$$
in terms of the new variables $C=1/\z$ and $\l^2=\o$, after a few rescalings.
The metric (14) reduces to that of eq.~(4) when using $\l^{-2}=e^{2\f}$. 

\section{Fivebrane instantons}

As was demonstrated in ref.~\cite{gs}, the BPS condition on the fivebrane 
instanton solution with the {\it vanishing} charges defines a gradient flow 
in the hypermultiplet moduli space. The flow implies the $SU(2)$ isometry 
of the UH metric since the non-degenerate action of this isometry in the 
four-dimensional UH moduli space gives rise to the well defined three-dimensional orbits 
that can be parametrized by the `radial' coordinate to be identified with the flow parameter. 

Let's consider a generic $SU(2)$-invariant metric in four Euclidean dimensions.
 In the Bianchi IX formalism, where the $SU(2)$ symmetry is manifest, the 
general {\it Ansatz} for such metrics reads \cite{tod3}
$$ ds^2=w_1w_2w_3dt^2+\fracmm{w_2w_3}{w_1}\s^2_1+\fracmm{w_3w_1}{w_2}\s^2_2+  
\fracmm{w_1w_2}{w_3}\s^2_3 \eqno(15)$$
in terms of the $su(2)$ (left)-invariant (Cartan) one-forms $\s_i$  and the
radial coordinate $t$. The metric (15) is dependent upon three functions
$w_i(t)$, $i=1,2,3$. 

Being applied to the metric (15), the ASD Weyl condition gives rise to a
 (Halphen) system of {\it Ordinary Differential Equations} (ODE) \cite{tod3,hit},
$$\eqalign{
 \dt{A}_1~=~&-A_2A_3 +A_1(A_2+A_3) ~,\cr
 \dt{A}_2~=~&-A_3A_1 +A_2(A_3+A_1) ~,\cr
 \dt{A}_3~=~&-A_1A_2 +A_3(A_1+A_2) ~,\cr}\eqno(16)$$
where the dots denote differentiation with respect to $t$, and the functions 
$A_i(t)$ are defined by the auxiliary ODE system,
$$\eqalign{
   \dt{w}_1~=~&-w_2w_3 +w_1(A_2+A_3)~,\cr
   \dt{w}_2~=~&-w_3w_1 +w_2(A_3+A_1)~,\cr
   \dt{w}_3~=~&-w_1w_2 +w_3(A_1+A_2)~.\cr}\eqno(17)$$

The Halphen system (16) has a long history. Perhaps, its most natural (manifestly integrable) derivation is 
provided via a reduction of the $SL(2,{\bf C})$ anti-self-dual Yang-Mills equations from four Euclidean 
dimensions to one. The Painlev\'e VI equation is known to be  behind the ASD-Weyl 
geometries having the $SU(2)$ symmetry \cite{tod3,hit}.  In fact, all quaternionic metrics 
with the $SU(2)$ symmetry are governed by the particular Painlev\'e VI equation:
$$\eqalign{
y''~=~&\fracmm{1}{2}\left( \fracmm{1}{y} 
+\fracmm{1}{y-1}+\fracmm{1}{y-x}\right)
(y')^2- \left( \fracmm{1}{x} +\fracmm{1}{x-1}+\fracmm{1}{y-x}\right)y'\cr
&~  +\fracmm{y(y-1)(y-x)}{x^2(x-1)^2}\left[
\fracmm{1}{8} -\fracmm{x}{8y^2}+\fracmm{x-1}{8(y-1)^2}
+\fracmm{3x(x-1)}{8(y-x)^2}\right] ~,\cr}\eqno(18)$$
where $y=y(x)$, and the primes denote differentiation with respect to $x$ \cite{hit}.

The equivalence between eqs.~(16) and (18) is well known to mathematicians \cite{tod3,hit}.
An exact solution to the Painlev\'e VI equation (18), which leads to a 
{\it regular} (and complete) quaternionic metric (15), is unique \cite{hit}.
The regular solution can be written down in terms of the standard 
theta-functions $\vq_{\a}(z|\t)$, where $\a=1,2,3,4$, and  the arguments are related as 
$z=\fracm{1}{2}(\t -k)$, where $k$ is an arbitrary (real and positive) parameter. 
The variable $\t$ is related to the variable $x$ of eq.~(18) via the relation
$$x=\vq^4_3(0)/\vq^4_4(0)~,\eqno(19)$$ 
where the value of the theta-function variable $z$ is explicitly indicated, as usual.
The explicit solution to eq.~(18) reads \cite{dub}
$$\eqalign{
y(x)~=~& \fracmm{\vq_1'''(0)}{3\p^2\vq^4_4(0)\vq_1'(0)}+\fracmm{1}{3}\left[
1+\fracmm{\vq_3^4(0)}{\vq^4_4(0)}\right] \cr
 &~ +\fracmm{\vq_1'''(z)\vq_1(z)-2\vq_1''(z)\vq_1'(z)+
2\p i(\vq_1''(z)\vq_1(z)-\vq_1'{}^2(z))}{2\p^2\vq_4^4(0)\vq_1(z)(\vq_1'(z)+
\p i\vq_1(z))}~~.\cr}\eqno(20)$$ 

The parameter $k>0$ describes the monodromy of this solution around 
its essential singularities (branch points) $x=0,1,\infty$. This (non-abelian)
 monodromy is generated by the matrices (with the purely imaginary eigenvalues
 $\pm i$) 
$$ M_1=\left( \begin{array}{cc} 0 & i \\ i & 0\end{array} \right)~,\quad
  M_2=\left( \begin{array}{cc} 0 & i^{1-k} \\ 
i^{1+k} & 0\end{array} \right)~,\quad
M_3=\left( \begin{array}{cc} 0 & i^{-k} \\ 
-i^{k} & 0\end{array} \right)~.\eqno(21)$$

The function (20) is meromorphic outside its essential singularities at 
$x=0,1,\infty$, while is also has simple poles at 
$\bar{x}_1,\bar{x}_2,\ldots$, where $\bar{x}_n\in (x_n,x_{n+1})$ and 
$x_n=x(ik/(2n-1))$ for each positive integer $n$. Accordingly, the metric is 
well-defined (complete) for $x\in (\bar{x}_n,x_{n+1}]$, i.e. inside the unit ball 
in ${\bf C}^2$ with the origin at $x=x_{n+1}$ and the boundary at $x=\bar{x}_n$ \cite{hit}.
Near the boundary the metric has the following asymptotical behaviour \cite{qum}:
$$\eqalign{
 ds^2~\propto~&\fracmm{dx^2}{(1-x)^2}+\fracmm{4}{(1-x)\cosh^2(\p k/2)}\s^2_1+
\fracmm{16}{(1-x)^2\sinh^2(\p k/2)\cosh^2(\p k/2)}\s^2_2\cr
&~ + \fracmm{4}{(1-x)\sinh^2(\p k/2)}\s^2_3~~+ ~~{\rm regular~terms}~.\cr}
\eqno(22)$$

As is clear from eq.~(22), the real parameter $k$ can be identified with the five-brane
instanton action that is proportional to the CY volume and $1/g^2_{\rm string}$ as well.
The semiclassical regime thus arises near the boundary $x\to 1^-$ at $k\to +\infty$. In
this limit, one gets back the Ferrara-Sabharwal metric out of that in eq.~(22) after rescaling
$(1-x)\to 2^6e^{\p k}(1-x)$ and redefining $x=r^2$. 

A few comments are in order.

The very notions of a `wrapped brane', an `instanton' and a `dilaton' are essentially semiclassical, and 
they do not exist non-perturbatively.  We consider the full UH theory as the NLSM, i.e. modulo field 
reparametrizations (or diffeomorphisms in the NLSM target space). The physical interpretation of the exact 
quaternionic solutions to the UH metric is, however, possible in the semiclassical regime. Hence, first, we 
identify the semilasssical region and, second, we rewrite a given exact solution as a sum of the known 
classical solution and the exponentially small corrections  with respect to the well defined real parameter 
(or modulus). Those corrections are finally identified with the instanton contributions, whose origin (due 
to the wrapped BPS branes) we already know in the context of the CY compactified type-IIA superstrings.

The supersymmetric 3-cycles ${\cc_3}$ are defined by two conditions: (i) the pullback of the CY K\"ahler
form $J$ on  ${\cc}_3$ should vanish, $\left. J\right|_{\cc_3}=0$, and (ii)  the pullback of the imaginary
part of the holomorphic CY 3-form $\O$ should vanish too,  $\left. {\rm Im}\,\O\right|_{\cc_3}=0$ \cite{bbs}.  
In the terminology of ref.~\cite{doug}, the supersymmetric 3-cycles $\cc_3$ are of the A-type, whereas the 
Calabi-Yau threefold itself is of the B-type. In our case, the only relevant modulus of a wrapped 5-brane is its
volume, i.e. the CY `size' parameter (a K\"ahler moduli). The semiclassical description can be valid only for
large CY volumes. In the opposite limit of small CY volumes, an exact N=2 superconformal field theory 
(Landau-Ginzburg) description can apply \cite{doug}. Mirror symmetry may allow us to relate these two different
descriptions. 

The coefficient at $\s^2_2$ in eq.~(22) vanishes faster than
the coefficients at $\s^2_1$ and $\s^2_3$ when approaching the boundary,
$x\to 1^-$. On the two-dimensional boundary annihilated by $\s_2$ one has the natural conformal 
structure
$$ \sinh^2(\p k/2)\s^2_1 +\cosh^2(\p k/2)\s^2_3~. \eqno(23)$$
The only relevant parameter $\tanh^2(\p k/2)$ in eq.~(23) represents the 
central charge (or the conformal anomaly) of the 2d conformal field theory on the boundary. Our results are,
therefore, consistent with the holographic principle \cite{last}.

I would like to thank Klaus Behrndt, Wolfgang Lerche, Nick Warner and Bernard de Wit for discussions. 
This work is supported in part by the `Deutsche Forschungsgemeinschaft'.

\end{document}